\documentclass{pasj00}

\begin{document}
\SetRunningHead{T. Kato et al.}{WZ Sge-Type Star V592 Her}

\Received{}
\Accepted{}

\title{WZ Sge-Type Star V592 Herculis}

\author{Taichi \textsc{Kato}, Makoto \textsc{Uemura}}
\affil{Department of Astronomy, Kyoto University,
       Sakyo-ku, Kyoto 606-8502}
\email{tkato@kusastro.kyoto-u.ac.jp}

\author{Katsura \textsc{Matsumoto}}
\affil{Graduate School of Natural Science and Technology, Okayama University,
       Okayama 700-8530}
\email{katsura@cc.okayama-u.ac.jp}

\author{Timo \textsc{Kinnunen}}
\affil{Sinirinnantie 16, SF-02660 Espoo, Finland}
\email{stars@personal.eunet.fi}

\author{Gordon \textsc{Garradd}}
\affil{PO Box 157, NSW 2340, Australia}
\email{loomberah@ozemail.com.au}

\author{Gianluca \textsc{Masi}}
\affil{Via Madonna de Loco, 47, 03023 Ceccano (FR), Italy}
\email{gian.masi@flashnet.it}

\email{\rm{and}}

\author{Hitoshi \textsc{Yamaoka}}
\affil{Faculty of Science, Kyushu University, Fukuoka 810-8560}
\email{yamaoka@rc.kyushu-u.ac.jp}



\maketitle

\begin{abstract}
   We observed the entire course of the 1998 outburst of V592 Her, which
was originally reported as a nova in 1968.  We have been able to construct
a full light curve of the outburst, which is characterized by a rapid
initial decline (0.98 mag d$^{-1}$), which smoothly developed into a
plateau phase with a slower linear decline.  We detected superhumps
characteristic to SU UMa-type dwarf novae $\sim$7 d after the optical
maximum.  The overall behavior of the light curve and the development of
superhumps were characteristic to a WZ Sge-type dwarf nova.  Combined
with the past literature, we have been able to uniquely determine the
superhump period to be 0.05648(2) d.  From this period, together with
a modern interpretation of the absolute magnitude of the outburst light
curve, we conclude that the overall picture of V592 Her is not inconsistent
with a lower main-sequence secondary star in contrast to a previous claim
that V592 Her contains a brown dwarf.
\end{abstract}

\section{Introduction}

   WZ Sge-type dwarf novae are a still enigmatic class of SU UMa-type
dwarf novae [for recent summaries of dwarf novae and SU UMa-type dwarf
novae, see \citet{osa96review} and \citet{war95suuma}, respectively],
which is characterized by a long ($\sim$ 10 yr) outburst
recurrence time and a large ($\sim$ 8 mag) outburst amplitude
(cf. \cite{bai79wzsge}; \cite{dow81wzsge}; \cite{pat81wzsge};
\cite{odo91wzsge}; \cite{kat01hvvir}).

   In recent years, the secondary stars (mass-donor stars) of WZ Sge-type
dwarf novae, or dwarf novae with extremely large outbursts amplitudes
(TOADs, \cite{how95TOAD}), have been regarded as promising candidates for
brown dwarfs (\cite{how97periodminimum}; \cite{pol98TOAD};
\cite{cia98CVIR}; \cite{pat01SH}; \cite{how01llandeferi}).
The existence of a brown-dwarf secondary star has been also considered to
play an important role in realizing an extremely low quiescent viscosity
of WZ Sge-type stars required (\cite{sma93wzsge}; \cite{osa95wzsge})
from the disk-instability theory (\cite{mey98wzsge};
\cite{min98wzsge})\footnote{
  Arguments, however, exist against the extremely low
  quiescent viscosity.  \citet{las95wzsge}, \citet{war96wzsge} assuming
  evaporation/truncation of the inner disk are the best examples.
  \citet{ham97wzsgemodel} and \citet{bua02suumamodel} presented slight
  modifications of these ideas.
  The discovery of a WZ Sge-type phenomenon in a long-period system
  \citep{ish01rzleo} suggests that the existence of a brown-dwarf
  secondary is not a necessary condition for manifestation of the
  WZ Sge-type phenomenon.
}
Observational confirmation of cataclysmic variables (CVs) with brown
dwarf secondaries is also important in that it can provide an independent
estimate of the upper limit of the age of the Universe (\cite{pol98TOAD};
\cite{szk02egcnchvvirHST}).
In particular, \cite{how01llandeferi} claimed the direct spectroscopic
detection of a brown dwarf in LL And, but inconsistency in this
interpretation has been later found \citep{how02llandefpegHST}.
\citet{men01j1050} reported a discovery of a CV with a brown dwarf
based on radial velocity studies.
V592 Her is another object in which the existence of a brown dwarf
has been claimed \citep{vantee99v592her}.

   V592 Her was discovered as a possible fast nova in 1968 on Sonneberg
plates \citep{ric68v592her}.  The observed maximum was $m_{\rm pg}$ = 12.3.
Although there was a gap in the 1968 observation, the outburst lasted
at least for 30 d.  \citet{ric68v592her} reported that the object was
exceptionally blue based on a comparison between quasi-simultaneously
taken blue- and yellow-sensitive plates.  Based on this conspicuously
blue color at maximum, \citet{due87novaatlas} suspected that the object
may be either a dwarf nova or an X-ray nova resembling V616 Mon.

   \citet{ric91v592her} further studied Sonneberg plates, and discovered
a second outburst in 1986.  The recorded maximum of the 1986 outburst
was $m_{\rm pg}$ = 13.6 (1986 May 12).  The limited coverage of this 1986
outburst made it difficult to draw a conclusion on the nature of this
outburst.  The star has been intensively monitored by visual observers,
members of the Variable Star Observers League in Japan (VSOLJ) since
1986 February.  The absence of visual detection of the 1986 May outburst
suggests that the outburst was fainter than the 1968 one, or the brightness
peak lasted for a very short time.  In spite of the intensive world-wide
efforts, no further outburst had been detected until 1998.

   On 1998 August 26.835 UT, Timo Kinnunen detected the object in outburst
at $m_{\rm v}$ = 12.0 (vsnet-alert 2067).\footnote{
$\langle$http://www.kusastro.kyoto-u.ac.jp/vsnet/Mail/\\alert2000/msg00067.html$\rangle$.}
He also noted a 0.5 mag variation within 0.08 d.
The last negative observation (fainter than 13.2) was made by Patrick Schmeer
on August 25.899 UT.  The object was reported to fade by 0.5--1.0 mag within
1 d of this detection.  Judging from the rapid fading and the presence of
an immediately preceding negative observation, this outburst must have been
caught around the peak brightness.  The dwarf nova-type nature was
subsequently confirmed with spectroscopy (Mennickent et al.,
vsnet-alert 2087.\footnote{
$\langle$http://www.kusastro.kyoto-u.ac.jp/vsnet/Mail/\\alert2000/msg00087.html$\rangle$.}; \cite{men02v592her}).
The large outburst amplitude ($>$ 9 mag, \cite{due87novaatlas};
\cite{ric91v592her}) and long recurrence times (10--20 yr) clearly qualifies
V592 Her as a best candidate for a WZ Sge-type dwarf nova.
\citet{due98v592her} reported
detection of superhumps, confirming that V592 Her belongs to SU UMa-type
dwarf novae.  \citet{vantee99v592her} argued, based on their quiescent
photometry, that the secondary star of V592 Her can be a brown dwarf.
We present a summary of our observations of the entire aspect of the 1998
outburst conducted as a part of VSNET Collaboration.\footnote{
$\langle$http://www.kusastro.kyoto-u.ac.jp/vsnet/$\rangle$.}

\section{Observations}

   The Kyoto observations were done using an unfiltered ST-7 camera
attached to a 25-cm Schmidt-Cassegrain
telescope.  The images were dark-subtracted,
flat-fielded, and analyzed using the Java$^{\rm TM}$-based PSF
photometry package developed by one of the authors (T. Kato).
The differential magnitudes of the variable were measured against
GSC 1518.1312 (GSC $V$ magnitude 11.62), whose constancy was confirmed
by comparison with GSC 1518.1287 (GSC $V$ = 11.74) and GSC 1518.1421
($V$ = 13.29, $B-V$ = +1.10).
The Loomberah observations were done using an unfiltered AP7 CCD camera
attached to a 45-cm f/5.4 Newtonian telescope.  The magnitudes of the
variable were measured using the same comparison as above, except on
September 22 when GSC 1518.756 ($V$ = 15.32, $B-V$ = +0.71) and
GSC 1518.662 ($V$ = 14.46, $B-V$ = +0.66) were used as the primary
comparison and check stars, respectively.
The Ceccano observations were done using an unfiltered ST-7 camera
attached to a 28-cm Schmidt-Cassegrain telescope.  The comparison stars
were the same as in the Kyoto observations.
The zero-point adjustments between the observations were made using
common comparison stars, Henden photometric sequence\footnote{
$\langle$ftp://ftp.nofs.navy.mil/pub/outgoing/aah/sequence/v592her.dat$\rangle$.
} and wide-field
CCD images taken on 1993 March 22 at Ouda Station \citep{Ouda}.
The resultant magnitudes were converted to a common system close to
R$_{\rm c}$, adopting $R_{\rm c}$ = 11.21 for GSC 1518.1312.
Since outbursting dwarf novae are known to have colors
close to $B-V$ = 0, the expected inaccuracy of zero-points caused by
different color responses of different CCDs will not affect the following
analysis.

   Barycentric corrections to
the observed times were applied before the following analysis.  The log
of observations is summarized in table \ref{tab:log}.

\begin{table*}
\caption{Journal of CCD photometry.}\label{tab:log}
\begin{center}
\begin{tabular}{crccrccc}
\hline\hline
\multicolumn{2}{c}{1998 Date}& Start--End$^*$ & Exp(s) & $N$
        & Mean mag$^\dagger$ & Error & Obs$^\ddagger$ \\
\hline
Aug. & 30 & 55.886--55.959 & 30 & 136 & 14.33 & 0.01 & G \\
     & 31 & 57.308--57.393 & 90 &  53 & 14.47 & 0.01 & M \\
Sep. &  1 & 57.888--57.893 & 30 &   8 & 14.51 & 0.02 & G \\
     &  2 & 58.868--58.969 & 30 & 194 & 14.68 & 0.01 & G \\
     &  3 & 59.859--59.980 & 30 & 236 & 14.73 & 0.01 & G \\
     &  6 & 62.889--62.965 & 30 & 126 & 15.21 & 0.01 & G \\
     &  6 & 63.288--63.351 & 90 &  38 & 15.13 & 0.02 & M \\
     &  7 & 63.865--63.950 & 30 & 149 & 15.25 & 0.01 & G \\
     &  8 & 64.999--65.063 & 30 &  73 & 15.42 & 0.11 & K \\
     &  9 & 65.921--66.048 & 30 & 228 & 15.48 & 0.04 & K \\
     & 10 & 66.924--67.026 & 30 & 158 & 15.60 & 0.03 & K \\
     & 11 & 67.939--68.042 & 30 &  41 & 15.74 & 0.12 & K \\
     & 12 & 68.919--69.036 & 30 & 235 & 15.65 & 0.04 & K \\
     & 13 & 69.896--69.913 & 30 &  31 & 15.71 & 0.01 & G \\
     & 13 & 69.914--70.038 & 30 & 213 & 15.83 & 0.05 & K \\
     & 15 & 71.950--71.972 & 30 &  35 & 15.76 & 0.06 & K \\
     & 16 & 72.877--72.937 & 30 &  98 & 15.96 & 0.01 & G \\
     & 16 & 72.930--73.036 & 30 & 110 & 15.78 & 0.05 & K \\
     & 20 & 76.922--77.041 & 30 & 195 & 19.18 & 0.46 & K \\
     & 22 & 78.881--78.907 & 175$^\S$ &  7 & 18.74 & 0.09 & G \\
Oct. &  2 & 88.911--88.990 & 30 & 153 & 20.71 & 1.99 & K \\
     &  3 & 89.927--90.006 & 30 &  82 & 17.86 & 0.68 & K \\
     &  4 & 90.917--90.994 & 30 & 152 & 19.02 & 0.66 & K \\
\hline
 \multicolumn{8}{l}{$^*$ BJD$-$2451000.} \\
 \multicolumn{8}{l}{$^\dagger$ System close to $R_{\rm c}$.} \\
 \multicolumn{8}{l}{$^\ddagger$ G (Garradd), M (Masi), K (Kyoto team).} \\
 \multicolumn{8}{l}{$^\S$ Each image is a stack of five 35 s exposures.} \\
\end{tabular}
\end{center}
\end{table*}

\section{Astrometry}

   An initial astrometric result from an outburst image was reported by
\citet{mas98v592heriauc}, who reported J2000.0 coordinates of
\timeform{16h 30m 56s.42}, \timeform{+21D 16' 58''.3}.  Since a discrepancy
from the coordinates measured from the 1968 outburst photograph suggested
a significant proper motion \citep{vantee99v592her}, we remeasured the
available CCD images on a modern astrometric grid.

   The resultant astrometry from the outburst CCD image by GM (epoch =
1998.657) is \timeform{16h 30m 56s.425}, \timeform{+21D 16' 58''.60}
(J2000.0, grid GSC-2.2.1, fitting error \timeform{0''.15}), which is
consistent with the value of \citet{mas98v592heriauc} (grid
USNO-A1.0).  From the DSS2 blue plate (epoch = 1994.420) we obtained
\timeform{16h 30m 56s.424}, \timeform{+21D 16' 58''.76} on the same
grid (fitting errors \timeform{0''.10}).  These values are almost
identical to the position of candidate star No. 1 in
\citet{due87novaatlas} (\timeform{16h 30m 56s.43}, \timeform{+21D 16'
58''.5}, precessed to J2000.0).  Other available plate scans
do not reveal this object with enough detail to perform astrometry.

   From our measurements only, the upper limit of the proper motion is
\timeform{0''.06} yr$^{-1}$, and Duerbeck's position (prior to 1986)
suggests that it would be much smaller.  On the other hand, the
position of V592 Her in quiescence reported by \citet{vantee99v592her}
(\timeform{16h 30m 56s.32} $\pm$ 0$^s$.04 , \timeform{+21D 16' 57''.9}
$\pm$ \timeform{0''.6}, epoch = 1997.589) is incompatible with these
values, especially in the Right Ascension.

   An inspection of POSS I blue plate scan (epoch = 1955.385) shows a
faint object about \timeform{5''} north of the above measured
position.  If it is really V592 Her, it seems to favor the presence of
a proper motion in the contrary direction to what was reported in 
\citet{vantee99v592her}.  We conclude that the claimed presence of a
high proper motion is still controversial and suggest that the measurements
of the 1968 outburst plates need to be reexamined using original plate
material.

\section{Result and Discussion}

\subsection{Overall Outburst Light Curve}\label{sec:lc}

\begin{figure*}
  \begin{center}
    \FigureFile(140mm,100mm){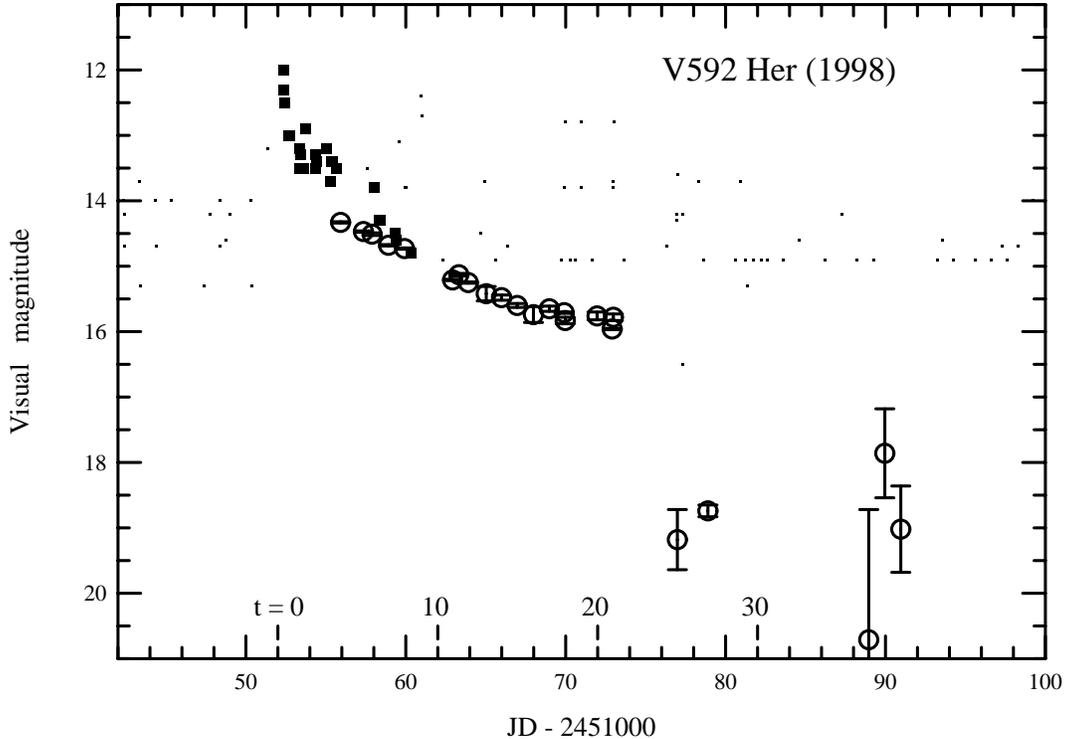}
  \end{center}
  \caption{Light curve of the 1988 superoutburst of V592 Her drawn from
  visual observations reported to VSNET.
  Large and small dots represent positive and negative (upper limit)
  observations, respectively.  Open circles with error bars represent
  nightly averaged CCD observations listed in table \ref{tab:log}.
  The epoch of maximum ($t$ = 0, see text) corresponds to JD 2451052.
  }
  \label{fig:lc}
\end{figure*}

   Figure \ref{fig:lc} shows the light curve of the 1988 superoutburst of
V592 Her drawn from visual observations reported to VSNET.  Large and small
dots represent positive and negative (upper limit) observations,
respectively.  Open circles with error bars represent nightly averaged
Kyoto CCD observations (table \ref{tab:log}).  The overall light curve is
characterized by the presence of a sharp maximum ($t$ = 0, JD 2451052
$\pm$1 d) followed by a rapid decay.  The decay became slower as the
object faded, and smoothly evolved into a gradually fading stage
(plateau phase).  Between $t$ = 21 and $t$ = 25, the object experienced
a sudden drop by 3.4$\pm$0.5 mag.  The early development of the light
curve more resembles those of very fast novae rather than those of
usual SU UMa-type dwarf novae, which are characterized by the presence
of a linear (exponential in flux scale) fading at a rate of
0.03--0.16 mag d$^{-1}$ (cf. \cite{war85suuma}; see also
\cite{kat02v359cen} for a summary of recent well-documented examples).
The sudden fading between $t$ = 21 and $t$ = 25 is characteristic of
the termination of a superoutburst in an SU UMa-type star.  The later
part of the outburst ($7 \leq t \leq $25) is more characteristic of
a usual SU UMa-type superoutburst while the initial part is more unusual.

   Similar departures of early light curves from the ``canonical"
light curve of SU UMa-type superoutbursts is rather commonly seen in
WZ Sge-type outbursts.  In WZ Sge itself (e.g. \cite{ort80wzsge};
\cite{pat81wzsge}), there seems to have been such a sharp initial
peak.\footnote{
  There exists an argument against the sharp, initial peak recorded
  in the past observations, because these feature were recorded on
  blue-sensitive photographs, which had a different sensitivity from
  visual observations.  See also \citet{kat01hvvir}.
}  Among well-observed WZ Sge-type outbursts, the present V592 Her
most clearly showed this feature.

   Such a deviation from a linear (exponential in flux scale) decay
in a WZ Sge-type outburst is shown to be naturally understood as
a consequence of a rapid viscous depletion
of the large amount of stored gas during the initial stage of a WZ Sge-type
outburst (\cite{osa95wzsge}; see also \cite{can93DI} for a basic model).
\citet{can01wzsge} recently successfully modeled the light curve of the 2001
superoutburst of WZ Sge with this mechanism.  \citet{can02ugem1985}
showed that this mechanism also worked in a system with a long orbital period.
The case of V592 Her is more striking than in the 2001 superoutburst of
WZ Sge.  A linear fit to the first 1 d of the light curve has yielded
a mean decay rate of 0.98 mag d$^{-1}$.  The decay rate decreased to
0.05 mag d$^{-1}$ during the late half of the plateau phase.  The initial
decay rate was 4 times larger than that of the 2001 superoutburst of
WZ Sge \citep{can01wzsge}.\footnote{
   The first 15-d average of the decline rate in V592 Her is comparable
   to that in WZ Sge \citep{can01wzsge}.  Since the effect of a viscous
   decay is stronger near the outburst peak, we use the initial decline
   rate described in the text.
}  By applying the relation between the
viscous decay time-scale ($\tau_\nu$) and the surface density in the disk
($\Sigma$), $\tau_\nu \propto \Sigma^{-3/7}$ \citep{can01wzsge} to the
initial decay rate, the initial surface density is estimated to be $\sim$25
times larger than that in the initial part (derived from an average of the
first 15 d) of the 2001 superoutburst of WZ Sge.

   WZ Sge-type dwarf novae are known to frequently (but not always) show
post-outburst rebrightenings [for a review, see \citet{kat98super}.
See also \citet{ric92wzsgedip}; \citet{how95TOAD}; \citet{kuu96TOAD};
\citet{kuu00wzsgeSXT}; \citet{kat97egcnc}; \citet{pat98egcnc}].\footnote{
  These phenomena are sometimes referred to as {\it echo outbursts},
but we avoid this terminology because this idea was first proposed
to describe the ``glitches" or ``reflares" in soft X-ray transients (SXTs)
\citep{aug93SXTecho}.  In SXTs, hard-soft transition is considered to
be more responsible for the initially claimed phenomenon
\citep{min96SXTtransition}, which is clearly physically different from
dwarf nova-type rebrightenings.
}  Due to the faintness of the object, the existence of such a
post-superoutburst rebrightening was not unambiguously confirmed during
the present outburst of V592 Her, although there may have been a hint
of such phenomenon on October 3 (JD 2451090).  However, a long-lasting
rebeightening as observed in AL Com (\cite{kat96alcom}; \cite{nog97alcom};
\cite{pat96alcom}), WZ Sge in 2001 (\cite{ish02wzsgeletter};
\cite{pat02wzsge}), and V2176 Cyg \citep{nov01v2176cyg} was not recorded.
There was no hint of a bright rebrightening as expected by
\citet{bua02suumamodel}.

\subsection{Superhump Period}

   \citet{due98v592her} reported the detection of superhumps (period
either 0.06007 d or 0.06391 d) based on their three-night observation.
A closer look at the data by \citet{due98v592her} left some uncertainty
regarding this period determination, mainly because only one superhump
per night was observed, which makes unique alias selection virtually
impossible.  In order to solve this problem, we have digitized the
figure in \citet{due98v592her} and measured their observations to an
accuracy of 0.001 mag and 0.0001$-$0.0002 d.  Although a period analysis
of these data has confirmed the claimed periods by \citet{due98v592her},
there remains substantial possibility around $P$ = 0.0567 d (see also
the upper panel of figure \ref{fig:pdm}).

\begin{figure}
  \begin{center}
    \FigureFile(88mm,120mm){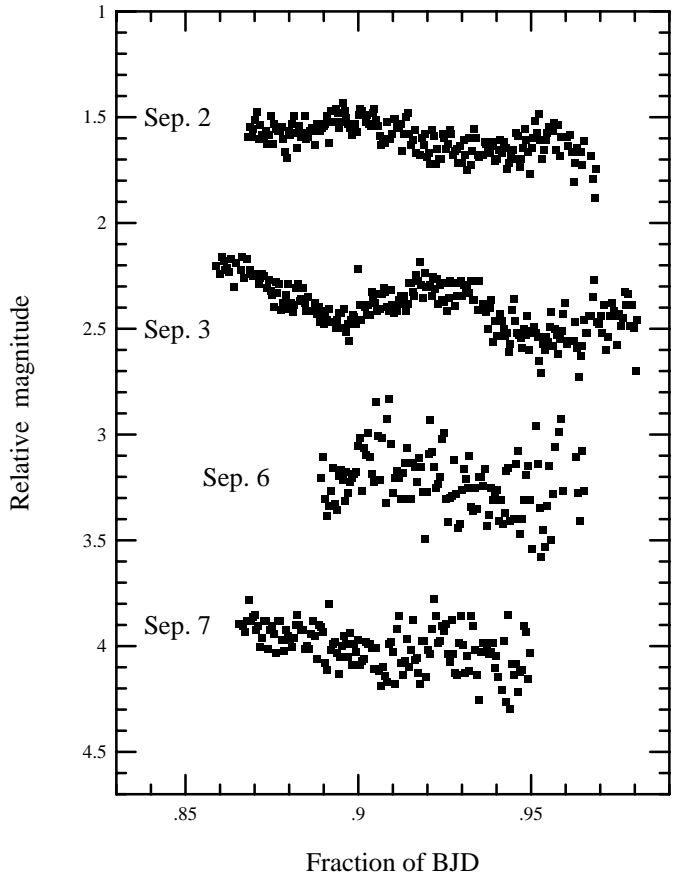}
  \end{center}
  \caption{Representative nightly superhump light curves during the
  superoutburst plateau.  Superhumps with amplitudes of 0.1--0.2 mag
  are present.  These observations covered an earlier epoch
  than in \citet{due98v592her}.
  }
  \label{fig:nightly}
\end{figure}

   We have further extracted the times of superhump maxima from our
observations between 1998 September 2 and September 7 (early part of the
superoutburst plateau, see figure \ref{fig:nightly}).
These observations covered an earlier epoch
than in \citet{due98v592her}.  The times of maxima were determined
by fitting the average superhump light curve (given in figure \ref{fig:shph})
to the observed data.  The maximum times and 1-$\sigma$ errors of timing
estimates were determined with Marquardt-Levenberg method \citep{Marquardt}.
The validity of the fits has also been confirmed with a comparison of
independent eye extraction of maximum times.
Table \ref{tab:sh} lists the measured timings of the superhump maxima.
The values are given to
0.0001 d in order to avoid the loss of significant digits in a later
analysis.  These maxima are not well expressed by either of the two
candidate periods listed in \citet{due98v592her}.
In particular, the interval of 0.396 d between the BJD 2451062.911 and
2451063.307 is only well expressed by a period near 0.057 d within their
respective errors (this interval corresponds to 6.59 and 6.20 cycles
of the two candidate periods \citep{due98v592her} of 0.06007 and 0.06391 d,
respectively).  We thus conclude that the short alias ($P \sim$ 0.0567 d)
is the true superhump period.  The cycle counts ($E$) in table \ref{tab:sh}
are calculated with this period.  A linear regression to the observed
superhump times gives the following ephemeris (the errors correspond to
1-$\sigma$ errors at the epoch of $E$ = 67) :

\begin{equation}
{\rm BJD (max)} = 2451058.9005(10) + 0.056498(13) E. \label{equ:reg1}
\end{equation}

\begin{table}
\caption{Timings of superhumps.}\label{tab:sh}
\begin{center}
\begin{tabular}{lrrrc}
\hline\hline
BJD$^*$ & Error$^\dagger$ & $E$$^\ddagger$ & $O-C_1$$^\dagger$$^\S$
             & Ref.$^\|$ \\
\hline
58.9003 & 10 &   0 &  $-$2 & 1 \\
58.9576 & 15 &   1 &     6 & 1 \\
59.8634 & 23 &  17 &    24 & 1 \\
59.9180 & 14 &  18 &     5 & 1 \\
62.9107 & 23 &  71 & $-$12 & 1 \\
63.3070 & 26 &  78 &  $-$3 & 1 \\
63.8718 & 25 &  88 &  $-$5 & 1 \\
63.9257 & 20 &  89 & $-$31 & 1 \\
64.4910 &  7 &  99 & $-$28 & 2 \\
65.5123 &  6 & 117 &    16 & 2 \\
67.4913 &  9 & 152 &    31 & 2 \\
\hline
 \multicolumn{5}{l}{$^*$ BJD$-$2451000.} \\
 \multicolumn{5}{l}{$^\dagger$ Unit 0.0001 d.} \\
 \multicolumn{5}{l}{$^\ddagger$ Cycle count.} \\
 \multicolumn{5}{l}{$^\S$ Against equation (\ref{equ:reg1}).} \\
 \multicolumn{5}{l}{$^\|$ 1: this work, 2: measured from} \\
 \multicolumn{5}{l}{\phantom{$^\|$} \citet{due98v592her}} \\
\end{tabular}
\end{center}
\end{table}

   Figure \ref{fig:pdm} shows the result of period analysis of superhumps
with the Phase Dispersion Minimization (PDM, \cite{PDM}).  The upper panel
shows an analysis of the data in \citet{due98v592her}, which shows the
possibility of many one-day aliases.  The lower panel shows an analysis
of the combined data (this work and \cite{due98v592her}), which covered
the superoutburst plateau between JD 2451057 (September 1) and 2451067
(September 11).  A strong preference of the frequency of 17.716(8) $d^{-1}$,
which corresponds to a period of $P$ = 0.05645(2) d, is clearly seen.
An exclusion of the data of \citet{due98v592her} did not significantly
change this trend.  The selection of the true alias is confirmed by these
analyses.
The significance level of this period is above 95 \%.  Figure \ref{fig:clean}
shows period analysis of superhumps in V592 Her with the Clean method
\citep{CLEAN}, with a gain parameter of 0.01.  The data and the frequency
range are the same as in the lower panel of figure \ref{fig:pdm}.
The Cleaned spectrum clearly shows that the frequency of 17.72 d$^{-1}$
is the only acceptable period.

   We finally adopted $P_{\rm SH}$ = 0.05648(2) from an average of
superhump timing analysis and PDM analysis.

\begin{figure}
  \begin{center}
    \FigureFile(88mm,120mm){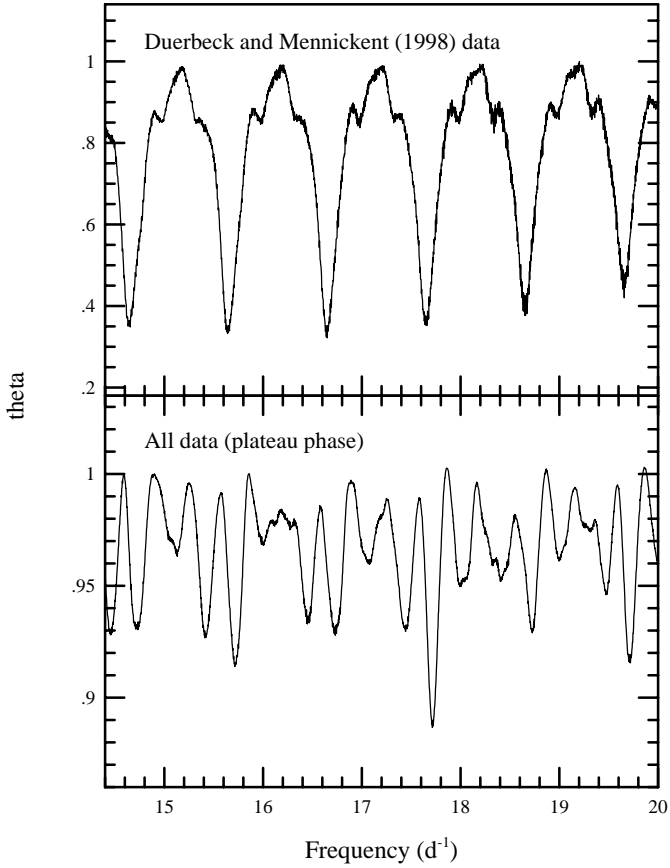}
  \end{center}
  \caption{Period analysis of superhumps in V592 Her with the Phase
  Dispersion Minimization (PDM).  (Upper) Analysis of the data in
  \citet{due98v592her}.  (Lower) Analysis of the data between JD
  2451057 (September 1) and 2451067 (September 11), which covered
  the superoutburst plateau.}
  \label{fig:pdm}
\end{figure}

\begin{figure}
  \begin{center}
  \end{center}
    \FigureFile(88mm,120mm){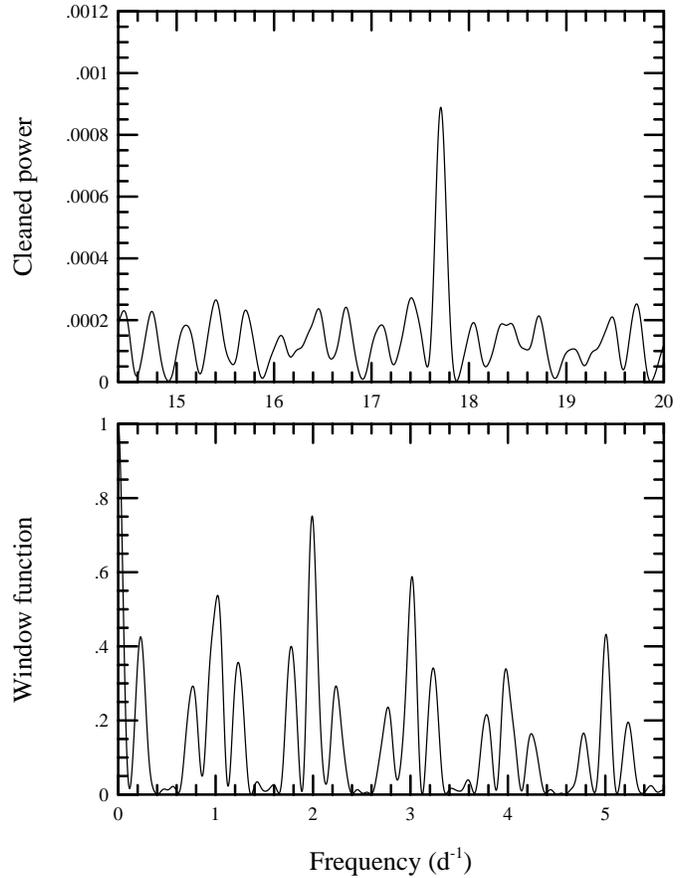}
  \caption{Period analysis of superhumps in V592 Her with the Clean
  method \citep{CLEAN}.  The data and the frequency range are the
  same as in the lower panel of figure \ref{fig:pdm}.
  (Upper) Cleaned spectrum (power in arbitrary
  unit). The frequency of 17.72 d$^{-1}$ is the only acceptable period.
  (Lower) Window function.
  }
  \label{fig:clean}
\end{figure}

   Figure \ref{fig:shph} shows a mean superhump profile phase-folded
with a period of $P_{\rm SH}$ = 0.05648 d.  The rapid rising and slowly
declining profile is characteristic to SU UMa-type superhumps
(\cite{vog80suumastars}, \cite{war85suuma}).  The mean amplitude
(0.15 mag) of superhumps is smaller than those of usual SU UMa-type
dwarf novae, but is close to that of a WZ Sge-type star, HV Vir
\citep{kat01hvvir}.

\begin{figure}
  \begin{center}
    \FigureFile(88mm,60mm){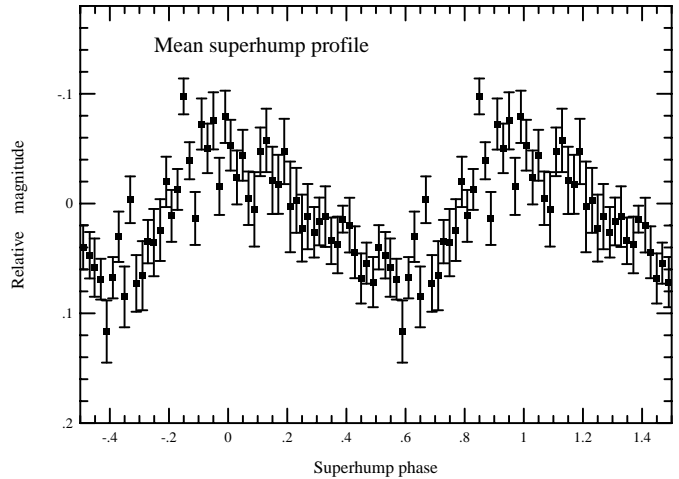}
  \end{center}
  \caption{Mean superhump profile of V592 Her phase-folded with a period
  of $P_{\rm SH}$ = 0.05648 d.}
  \label{fig:shph}
\end{figure}

   The newly established superhump period ($P_{\rm SH}$ = 0.05648(2) d)
is extremely close to those of WZ Sge ($P_{\rm SH}$ = 0.05726(1) d:
\cite{ish02wzsgeletter}, \cite{pat02wzsge}),
AL Com ($P_{\rm SH}$ = 0.05722(1) d: \cite{kat96alcom}, \cite{pat96alcom}),
the two best-studied WZ Sge-type dwarf novae.  All known WZ Sge-type
dwarf novae have $P_{\rm SH}$ shorter than 0.060 d except RZ Leo and EG Cnc
(see e.g. \cite{kat01hvvir}).  Among them, the long period of RZ Leo is
compatible with the evidence of a relatively massive secondary
\citep{ish01rzleo}.  Since the secondary of V592 Her is apparently
less luminous \citep{vantee99v592her} than in RZ Leo, the new period
better fits the general WZ Sge-type characteristics without necessarily
introducing, as we will see, a possibility of a brown dwarf secondary.

\subsection{Superhump Period Change}

\begin{figure}
  \begin{center}
    \FigureFile(88mm,60mm){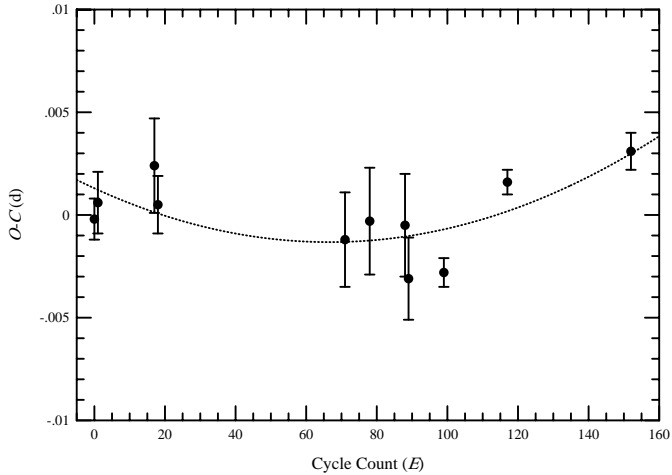}
  \end{center}
  \caption{$O-C$ diagram of superhump maxima.
  The parabolic fit corresponds to $\dot{P}$ = 1.2 $\pm \times$ 10$^{-6}$
  cycle$^{-1}$.}
  \label{fig:oc}
\end{figure}

   In contrast to the ``textbook" decrease of the superhump periods in
usual SU UMa-type dwarf novae (e.g. \cite{war85suuma}; \cite{pat93vyaqr}),
WZ Sge-type dwarf novae are recently known to show virtually zero or even
increase of the superhump periods (for a summary, see \cite{kat01hvvir}).
The quadratic term determined from the superhump maximum timings
corresponds to $\dot{P}$ = +1.2 $\pm$ 0.4 $\times$ 10$^{-6}$ cycle$^{-1}$
or $\dot{P}/P$ = +2.1(0.8) $\times$ 10$^{-5}$.  This value indicates
a small, but significant, period increase in V592 Her (figure \ref{fig:oc}).
This rate is comparable to the period changes observed in WZ Sge
(\cite{ish02wzsgeletter}; \cite{pat02wzsge}).

\subsection{Early Superhumps and Orbital Period}\label{sec:ESH}

   All well-observed WZ Sge-type dwarf novae are known to show double-humped
modulations having a period very close to the system orbital period
during the earliest stage of superoutbursts (\cite{kat96alcom};
\cite{mat98egcnc}; \cite{pat96alcom}; \cite{nog97alcom}; \cite{ish01rzleo};
\cite{kat01hvvir}; \cite{ish02wzsgeletter}; \cite{pat02wzsge}).
These modulations are called early superhumps.\footnote{
  This feature is also referred to as {\it orbital superhumps}
  \citep{kat96alcom}, {\it outburst orbital hump} \citep{pat98egcnc}
  or {\it early humps} \citep{osa02wzsgehump}.
}  The presence of early superhumps is the unique characteristic of
WZ Sge-type dwarf novae \citep{kat01wxcet}.
Although the origin of early superhumps is controversial, several
interpretations have been historically proposed: (1) enhanced hot spot
caused by a sudden increase of the mass-transfer (\cite{pat81wzsge};
\cite{pat02wzsge}), (2) immature form of superhumps \citep{kat96alcom},
(3) geometrical effect of a jet or a thickened edge of the accretion
disk \citep{nog97alcom}.

   Most recently, \citet{osa02wzsgehump} proposed that these humps are a
manifestation of a tidal 2:1 resonance in the accretion disks of binary
systems with extremely low mass ratios.  During the 2001 superoutburst
of WZ Sge \citep{ish02wzsgeletter}, a two-armed spiral velocity pattern
in the Doppler tomograms of the He\textsc{II} line was found
(\cite{ste01wzsgeiauc7675}; \cite{bab02wzsgeletter}) at the same time
of the appearance of early superhumps.  \citet{kat02wzsgeESH} suggested
that both early superhumps and the two-armed spiral velocity pattern
can be naturally considered by taking into the effect of a velocity
field of a tidally distorted disk (\cite{sma01tidal}; \cite{ogi02tidal}).

   In the present case of V592 Her, the apparent presence of early epoch
short-term variation (up to 0.5 mag) as inferred from visual observations
seems to suggest the presence of early superhumps as in the 2001 outburst
of WZ Sge.  However, the lack of time-resolved photometry during the
earliest stage of the outburst makes it difficult to draw a firm conclusion.

   By using the best-established fractional superhump excesses
($\epsilon=P_{\rm SH}/P_{\rm orb}-1$) of 1.0$\pm$0.1 \% in WZ Sge
(\cite{ish02wzsgeletter}; \cite{pat02wzsge}) and AL Com
(\cite{kat96alcom}; \cite{nog97alcom}; \cite{pat96alcom}), the orbital
period ($\sim$ period of early superhumps) is expected to be 0.05592(3) d.

   Most recently, \citet{men02v592her} reported candidate orbital periods of
91.2 $\pm$ 0.6 min ($P_1$: 0.0633(4) d),
85.5 $\pm$ 0.4 min ($P_2$: 0.0594(3) d) or
80.8 $\pm$ 0.6 min ($P_3$: 0.0561(4) d) from an analysis of their
spectroscopy taken a fews days after the maximum of the 1998 outburst.
Based on our identification of the true $P_{\rm SH}$, $P_3$ is now confirmed
to be the true $P_{\rm orb}$.  This difference of preferable period
selection between \citet{men02v592her} and this work can be reasonably
attributed to a severe aliasing clearly seen in the Fig. 3 of
\citet{men02v592her}.  Our period and $P_3$ in \citet{men02v592her}
are consistent within their respective errors.

\begin{figure}
  \begin{center}
    \FigureFile(88mm,60mm){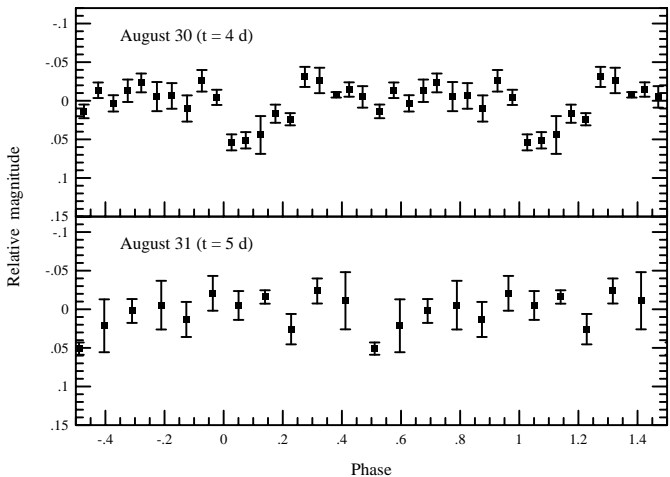}
  \end{center}
  \caption{Light curves on 1998 August 30 and 31 phase-averaged with the
  expected ($P$ = 0.05592 d) period of early superhumps.  The phase zero
  was arbitrarily taken as BJD 2451000.  On August 30, there is a hint
  of low-amplitude double-wave modulation, resembling early superhumps in
  HV Vir \citep{kat01hvvir}.  On August 31, only small-amplitude
  variation seems to have been marginally detected.}
  \label{fig:esh}
\end{figure}

Figure \ref{fig:esh} shows the light curves on 1998 August 30 and 31
($t$ = 4 d and 5 d, respectively) phase-averaged with a period of
0.05592 d, assuming that these variations reflect
early superhumps.\footnote{
   Due to the shortness of each runs, any trial period between 0.05592 d
   (adopted $P_{\rm orb}$) and 0.05648 d ($P_{\rm SH}$) gives the virtually
   same waveform.  Strictly speaking, we cannot distinguish early
   superhumps from (the growing stage of) superhumps from these observation
   only.  However, we consider it likely that these modulations reflect early
   superhumps because the transition from early superhumps to superhumps
   has been to confirmed to occur less than 1 d in WZ Sge
   (\cite{ish02wzsgeletter}; \cite{pat02wzsge}).  A chance to observe
   the growing stage of superhumps on two nights is expected to be very
   small.
}
On August 30, there is a hint of low-amplitude double-wave modulation
(with a rather strong signature of minimum), resembling early superhumps
in HV Vir \citep{kat01hvvir}.  On August 31, only small-amplitude
variation seems to have been marginally detected.  The amplitude was less
than 0.08 mag on August 31.  The weakness of the signal on August 31
has made it impossible to make a period determination from these
observations.

   Since both \citet{osa02wzsgehump} and \citet{kat02wzsgeESH} imply
that the amplitude of early superhumps is a strong function of the binary
inclination, the low amplitude of the possible early superhumps suggests
a low binary inclination.  This suggestion is consistent with
the lack of a strong He\textsc{II} emission line \citep{men02v592her},
which was strongly seen in emission in the
high-inclination system WZ Sge (\cite{ste01wzsgeiauc7675};
\citet{bab02wzsgeletter}), and the lack of large-amplitude variability in
quiescence \citep{vantee99v592her}.

\subsection{Brown Dwarf Secondary?}

   V592 Her is proposed to have a brown dwarf secondary because a main
sequence secondary would imply a large distance, which is inconsistent
with absolute magnitudes of ordinary dwarf novae
(\cite{vantee99v592her}).  Here we reexamine the distance and the
nature of the secondary based on the more accurate orbital period of
V592 Her.

   We can estimate a distance of a dwarf nova from the observed peak
magnitude and the empirically expected absolute magnitude.  The
correlation between the peak magnitude ($M_V({\rm max})$) and
the orbital period ($P_{\rm orb}$) provides $M_V({\rm max})\sim$
3.8--5.3 for a superoutburst (\cite{war95book};
\cite{vantee99v592her}).  When we apply this method to WZ Sge stars,
however, the observed peak magnitude is not suitable to estimate a
distance because they experience a rapid fading phase just after the
peak, during which the viscosity decays with time (\cite{can01wzsge};
\cite{osa95wzsge}).  After this viscosity decay phase, the surface density
of the accretion disk is expected to follow the same time-evolution
as in superoutbursts of ordinary SU UMa-type dwarf novae
\citep{osa95wzsge}.

   The peak magnitudes of these ordinary outbursts
are limited by the critical surface density ($\Sigma_{\rm max}$) required
by the disk instability theory \citep{can98DNabsmag}.  The calculated
peak magnitudes are known to well reproduce the Warner's relation.
\citet{can98DNabsmag} originally restricted the discussion to SS Cyg-type
dwarf novae, but the same discussion can be naturally extended to the upper
limits of $M_V({\rm max})$ of normal outbursts of SU UMa-type dwarf novae.
Superoutbursts are generally $\sim$0.5 mag brighter than upper-limit
magnitudes of normal outbursts \citep{war95suuma}, caused by an extra
heating by tidal dissipation, a safe upper limit of $M_V({\rm max})$
for ordinary SU UMa-type superoutbursts is estimated
to be 0.5 mag brighter than the extrapolation of \citet{can98DNabsmag}.
In the case of V592 Her, this value corresponds to $M_V({\rm max})
\sim +4.8$, which is consistent with the reported $M_V({\rm max})\sim$
3.8--5.3 for observed superoutbursts (\cite{war95book}).

   We should hence compare the expected $M_V({\rm max})$ not with the
observed peak magnitude, but with the magnitude at
which the viscous decay finishes, in other words, an ordinary plateau
phase begins.  As can be seen in figure \ref{fig:lc} and a comparison
with a simulation in \citep{osa95wzsge}, V592 Her experienced this phase
change at $V=14.0$ (considering that CCD observations tended to give
slightly fainter magnitudes than visual observations, and considering
the difficulty in accurately estimating such a faint magnitude visually,
this magnitude would better be regarded as an upper limit of the plateau
phase).  While \citet{vantee99v592her} estimate a distance
$d\sim$ 220--440 pc using a peak magnitude of $V=12$, our
estimation hence provides larger distances of $d\geq$ 550--1100 pc.

   Another caveat in \citet{vantee99v592her} is that they used a wrong
(longer) superhump period based on \citet{due98v592her}.  By adopting
the correct $P_{\rm SH}$ = 0.05648(2) and estimated $P_{\rm orb}$ =
0.05592(3) d, the expected absolute magnitude of a main-sequence secondary
filling the Roche-lobe of this $P_{\rm orb}$ is at least $\sim$1.0 mag
fainter \citep{bar98lowmassstars}.  Based on the same method of estimate
in \citet{vantee99v592her}, the
lower limit of the distance from a comparison of apparent magnitude
and the absolute magnitude of a main-sequence secondary is now lowered
to 900 pc or even lower.  This lower limit of the distance is now
not at all inconsistent with an estimate from the outburst photometry.

   The present new determinations of the true $P_{\rm SH}$, $P_{\rm orb}$
and the new distance estimate are thus consistent with a lower
main-sequence secondary.  By considering a main-sequence secondary
with $M_I$ = 12.4 (which corresponds to an upper limit of the luminosity
of a main sequence filling the Roche-lobe of V592 Her), the observed
color $R-I \sim$ 0.2 can be naturally explained by a contribution of
this secondary star.  In this case, we don't need to assume an extremely
cold ($\sim$10000 K) white dwarf as deduced in \citet{vantee99v592her}.
Although accurate determination of the white dwarf temperature should
await optical-UV spectroscopy, this finding seems to be consistent with
recent determinations of white dwarf temperatures in WZ Sge-type
dwarf novae (EG Cnc: 11700--13000 K, HV Vir: 12500--14000 K
\cite{szk02egcnchvvirHST}; GW Lib: 14700 K in average \cite{szk02gwlibHST};
LL And: 15000 K \cite{how02llandefpegHST}).  The present conclusion is
also comparable to recent result in WZ Sge itself
\citep{ste01wzsgesecondary}, who concluded that a lower main-sequence
secondary is still viable in spite of all the past negative efforts in
directly detecting a signature of emission from the secondary of WZ Sge.
In conjunction with the present conclusion, the presence of a brown dwarf
in WZ Sge-type dwarf novae is still an open question even in most promising
cases.\footnote{
  \citet{men02v592her} concluded, from their identification of $P_{\rm orb}$,
  that $\epsilon$ supports the brown-dwarf like nature of the secondary
  star.  However, as discussed in subsection \ref{sec:ESH}, the correct
  $P_{\rm orb}$ is their $P_3$ = 0.0561(4) d.  This value gives
  $\epsilon$ = 0.7 $\pm$ 0.7\%, which essentially gives no constraints
  on the existence of a brown dwarf.
  Furthermore, velocity fields of emission lines in WZ Sge-type superoutbursts
  are known to be very complex \citep{bab02wzsgeletter}, or even
  systematically vary \citep{kat02wzsgeESH}.  Radial velocity variation of
  emission lines in WZ Sge-type superoutbursts thus may not reasonably trace
  the binary motion as {\it a priori} assumed in \citet{men02v592her}.
}

\subsection{Related Objects}

   As stressed in \citet{kat01hvvir}, light curves of some WZ Sge-type
dwarf novae often display similar characteristics to those of very fast
novae.  The present, first-ever fully obtained, light curve of V592 Her
(figure \ref{fig:lc}) marks an even stronger similarity.  In WZ Sge
itself, either a lower surface density at the beginning of an outburst,
or a self-shielding effect arising from a nearly edge-on view, may have
reduced this effect.  In this context, the present light curve of
V592 Her even ``better" reproduce the expected light curve of a WZ Sge-type
dwarf nova \citep{osa95wzsge}.  As seen in subsection \ref{sec:ESH},
this light curve may be a result of a low binary inclination in V592 Her.
Among the stars listed in \citet{kat01hvvir}, V358 Lyr \citep{ric86v358lyr},
LS And \citep{sha78lsand} and
V4338 Sgr have very similar light curves to that of V592 Her.  These
objects may comprise a group of WZ Sge-type dwarf novae which is either
characterized by a stronger effect of initial viscous decay, or a low
binary inclination.  None of these systems, including V592 Her
\citep{vantee99v592her}, have been detected in ROSAT surveys
(\cite{ver97ROSAT}; \cite{ROSATRXP}).  Apparently low X-ray luminosities
of these systems makes a striking difference from the relatively
strong quiescent X-ray detection in WZ Sge \citep{ver97ROSAT}.  This
difference from WZ Sge may be a result of an even smaller quiescent
viscosity, which could explain a stronger effect of initial viscous decay.

\section{Summary}

   We observed the entire course of the 1998 outburst of V592 Her, which
was originally reported as a nova in 1968.  We have been able to first time
construct a full light curve of the outburst, which is characterized
by a rapid initial decline (0.98 mag d$^{-1}$), which smoothly developed
into a plateau phase with a slower linear decline.  The initial rapid
decay has been interpreted as a result of viscous decay theoretically
and naturally expected for a high surface-density accretion disk in
a WZ Sge-type outburst.
   We detected superhumps characteristic to SU UMa-type dwarf novae
$\sim$7 d after the optical maximum.  The overall behavior of the light
curve and the development of superhumps were characteristic to
a WZ Sge-type dwarf nova, although there was little evidence of early
superhumps, which may have been either escaped from detection because
of the unfavorable observational coverage or because of a low orbital
inclination.
   We examined astrometry of V592 Her using modern material, and have
yielded a safe upper limit (\timeform{0''.06} yr$^{-1}$) of its proper
motion.  The result in \citet{vantee99v592her} would have been somehow
overestimated.
   Combined with the past literature, we have been able to uniquely determine
the superhump period to be 0.05648(2) d.  We detected a small, but
significant, positive period change ($\dot{P}/P$ = +2.1(0.8) $\times$
10$^{-5}$) of superhumps.  We estimated an expected orbital
period of 0.05592(3) d.  From these periods, together with the modern
interpretation of the absolute magnitude of the outburst light curve,
we conclude that the overall picture of V592 Her is not inconsistent
with a lower main-sequence secondary star in contrast to a previous claim
that V592 Her contains a brown dwarf.

\vskip 3mm

We are grateful to many VSNET observers who have reported vital observations.
This work is partly supported by a grant-in aid [13640239 (T. Kato),
14740131 (HY)] from the Japanese Ministry of Education, Culture, Sports,
Science and Technology.
Part of this work is supported by a Research Fellowship of the
Japan Society for the Promotion of Science for Young Scientists (KM, MU).
The CCD operation at Loomberah is supported by The Planetary Society Gene
Shoemaker NEO Grant.
GM acknowledges the support by Software Bisque and Santa Barbara Instrument
Group.
This research has made use of the Digitized Sky Survey producted by STScI, 
the ESO Skycat tool, and the VizieR catalogue access tool.
This research has made use of the USNOFS Image and Catalogue Archive
operated by the United States Naval Observatory, Flagstaff Station
(http://www.nofs.navy.mil/data/fchpix/).

\end{document}